\documentclass[sigconf,10pt]{acmart}
\usepackage{array,tabularx,placeins,xurl,stfloats}
\graphicspath{{images/}}
\setcopyright{none}
\acmConference[MobiSys '27]{MobiSys 2027}{}{}
\acmYear{2027}
\copyrightyear{2027}
\acmDOI{}
\acmISBN{}

\title{AgentPRO: Coordinating Compiler Optimizations for LLM Agent Workflows}
\author{Jie Ren}
\affiliation{
  \institution{Shaanxi Normal University}
  \city{Xi'an}
  \country{China}
}

\author{Zheng Wang}
\affiliation{
  \institution{University of Leeds}
  \city{Leeds}
  \country{United Kingdom}
}

\begin{document}
\begin{abstract}
LLM agents repeatedly return to a model to generate actions and
interpret tool results, incurring repeated context processing and
generation overhead. Yet a value needed to execute an action need
not be observed by the model to generate that action. This distinction
creates an optimization opportunity: reconsider where model decisions
are necessary, together with the work performed between them.

We present AgentPRO, a compiler framework for optimizing agent
workflows as execution information becomes available. AgentPRO
organizes planned duties, generation regions, and concrete operations
in a hierarchical representation, distinguishing execution dependencies
from observations required for generation. Region transformations
combine compatible duties into fewer model requests, while operation
transformations reduce redundant context and tool work. These scopes
interact: eliminating a redundant read can remove an observation
requirement that otherwise separates generation regions. AgentPRO
coordinates transformations using initial plans, current observations,
and historical execution profiles through just-in-time, adaptive,
and profile-guided optimization. Runtime guards check state-dependent
assumptions and retain fallback paths.

We implement guarded workflow transformations for AppWorld agents.
On 115 tasks selected by successful historical baseline executions,
automatic AgentPRO reduces all-attempt mean token consumption by
41.85\% and mean component time by 6.61\% relative to those baselines,
successfully completing 110 of 115 tasks.

\end{abstract}

\maketitle

\section{Introduction}
\label{sec:introduction}

LLM agents translate natural-language requests into model decisions and
tool actions, supporting tasks from information retrieval to content
recommendation. This convenience can come with substantial execution cost,
even for a short request. On a movie-recommendation benchmark---selecting
one of four candidate films based on four reference films---LLMCompiler
reports approximately 20,000 input tokens and 230 output tokens for its
ReAct baseline~\cite{llmcompiler}. Processing inputs across repeated model
calls can therefore dominate the token cost of a seemingly modest task.

\begin{figure*}[!tbp]
  \centering
  \includegraphics[width=\linewidth,height=0.75\textheight,keepaspectratio]{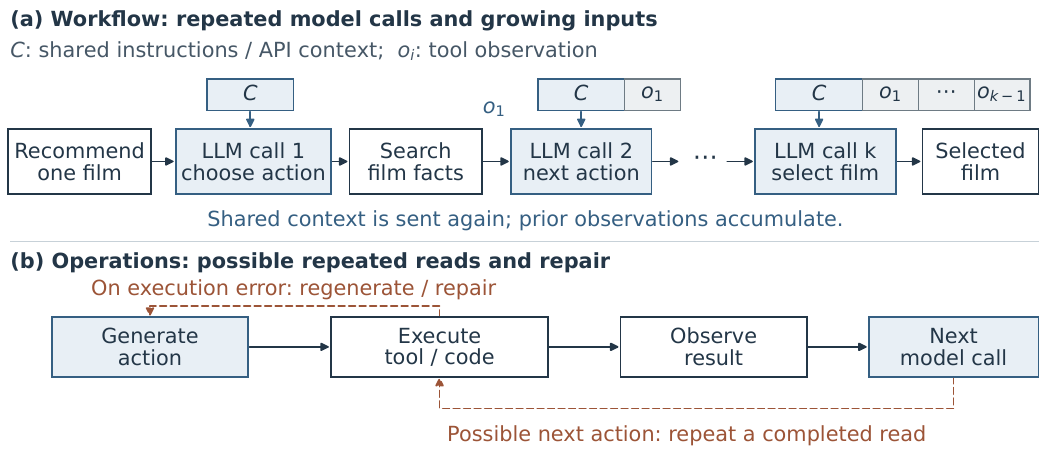}
  \caption{Schematic sources of agent token consumption. (a) Successive model
  calls receive shared context $C$ and accumulating tool observations $o_i$.
  (b) Repeated reads and execution errors can introduce additional model and
  tool work. Dashed paths illustrate possible behavior; they do not
  reconstruct the measured movie-recommendation benchmark trace.}
  \label{fig:intro-token-cost-workflow}
\end{figure*}

The source of this cost lies partly in how agent execution is organized.
Each model call can receive the same instructions and API descriptions,
together with an expanding history of tool observations
(Figure~\ref{fig:intro-token-cost-workflow}(a)). AgentDiet identifies
irrelevant, redundant, and expired information in these trajectories and
reduces it during execution~\cite{agentdiet}. Generated actions can repeat
completed reads, and execution errors can trigger further generation and
repair (Figure~\ref{fig:intro-token-cost-workflow}(b)). Reducing these costs
requires deciding what each call must generate, which context it needs,
and which tool operations remain necessary. These decisions span both the
boundaries between model calls and the work performed within them.

Compiler techniques make such execution choices explicit and optimizable.
LLMCompiler extracts parallelism among function calls~\cite{llmcompiler}.
SkVM adapts skills to model capabilities and combines compilation with
runtime code solidification and adaptive recompilation~\cite{skvm}.
ACON refines context-compression guidelines through agent failure
analysis~\cite{acon}, while TraceCompiler converts execution traces into
mostly deterministic workflows with residual model decisions~\cite{tracecompiler}.
These efforts use program structure, runtime state, or execution history
to guide optimization. Building on this direction, we study how to coordinate
transformations of generation structure and concrete operations as an agent
workflow unfolds.
Two properties of agent execution make this coordination challenging.

First, optimizations at different scopes are coupled. A later stage may
request a record that an earlier tool call has already returned. Reusing
that record can eliminate a read; if the read was the only intervening
observation requirement, it can also enable neighboring duties to be
generated together. Conversely, regrouping duties changes the context and
work assigned to the model. Region and operation transformations therefore
need a shared account of dependencies, observations, and effects to establish
when they can be applied together.

Second, the information needed for these decisions arrives at different
times. An initial workflow exposes intended duties, but the model determines
concrete actions during execution. Tool results reveal available values and
completed work, while traces from earlier runs reveal recurring patterns.
A plan alone cannot resolve every opportunity, and historical similarity
cannot establish the validity of a current result. Moreover, a legal
transformation need not be profitable: a larger generation region may
increase model difficulty and recovery cost. Optimization must track both
the remaining computation and the evidence available to transform it.

We present \textbf{AgentPRO}, a hierarchical compiler and runtime framework
for reducing agent token consumption and latency. Its central idea is to
optimize a shared representation as execution knowledge develops, using
initial structure, current observations, and historical profiles at their
respective decision points. This connects two dimensions of optimization:
\emph{where} to transform the workflow and \emph{when} sufficient information
is available to do so.

AgentPRO represents workflow structure, generation regions, and concrete
operations in one hierarchical intermediate representation (IR). The
prototype links planned stages to AST views of generated code and records
the contracts, observations, and completion requirements available to its
guarded rewrites. Region passes group or split generation duties; operation
passes change the context and actions within those regions. This shared
view makes dependencies between region and operation rewrites explicit,
while retaining unresolved requirements.

Three mechanisms apply this pass library throughout execution and across
runs. \emph{JIT} optimizes the initial workflow using its structure and
available contracts. \emph{Adaptive optimization} uses generated operations
and current observations to revise unfinished work. \emph{Profile-guided
optimization (PGO)} uses prior execution traces to choose transformations
for subsequent runs, including historically useful grouping and recurring
adaptive rewrites. These mechanisms share transformations; they differ in
the information that guides their application. Runtime checks validate
state-dependent assumptions and retain fallback paths when those assumptions
fail.

A policy controller chooses which mechanisms to enable, and a bounded,
rule-based pass manager applies eligible transformations to unfinished work.
Completed work supplies evidence for these decisions. Explicit optimization
and recovery accounting make the cost of each configuration visible.
Cost-guided pass selection and selective revisiting are extensions of this
design, distinct from the evaluated task-level configuration policies.

We implement AgentPRO for AppWorld plan-and-execute agents~\cite{appworld}
and evaluate it with DeepSeek on 115 tasks selected by successful stored
baseline executions. Automatic AgentPRO reduces all-attempt mean token
consumption by 41.85\% and mean component time by 6.61\% relative to that
historical baseline, completing 110 of 115 tasks. These headline results use
one recorded attempt per task and retain failed executions and recovery.
Separate studies compare fixed JIT, adaptive, and PGO combinations and
evaluate a learned selector with scenarios held out from training. The
configuration studies show task-dependent trade-offs; the learned selector
does not outperform its training-selected fixed comparator. Profile
acquisition is accounted for separately.

This work contributes:
\begin{itemize}
  \item \textbf{A hierarchical compiler framework for agent workflows.} AgentPRO represents workflows, generation regions, and concrete operations in a shared hierarchical IR, enabling region and operation passes to cooperate under explicit dependencies and validity conditions.
  \item \textbf{Unified lifecycle optimization and policy selection.} AgentPRO coordinates JIT, adaptive optimization, and PGO using plan structure, current observations, and historical profiles. Task-level policies choose optimization configurations, while a shared pass manager applies eligible guarded transformations.
  \item \textbf{Implementation and systematic evaluation.} We implement AgentPRO and evaluate its configurations on 115 baseline-qualified AppWorld tasks, measuring tokens, latency, success, optimization overhead, and recovery costs to characterize their benefits and trade-offs.
\end{itemize}

\section{Motivation: Optimizing the Plan and Its Execution}
\label{sec:motivation}

An agent pays to generate actions, supply context, and execute operations.
The plan exposes intended work; the generated program reveals concrete
actions; previous runs record useful choices. Can an optimizer use each
source when it becomes available, while preserving the task's requirements?

\begin{figure*}[!tbp]
  \centering
  \includegraphics[width=\linewidth,height=0.75\textheight,keepaspectratio]{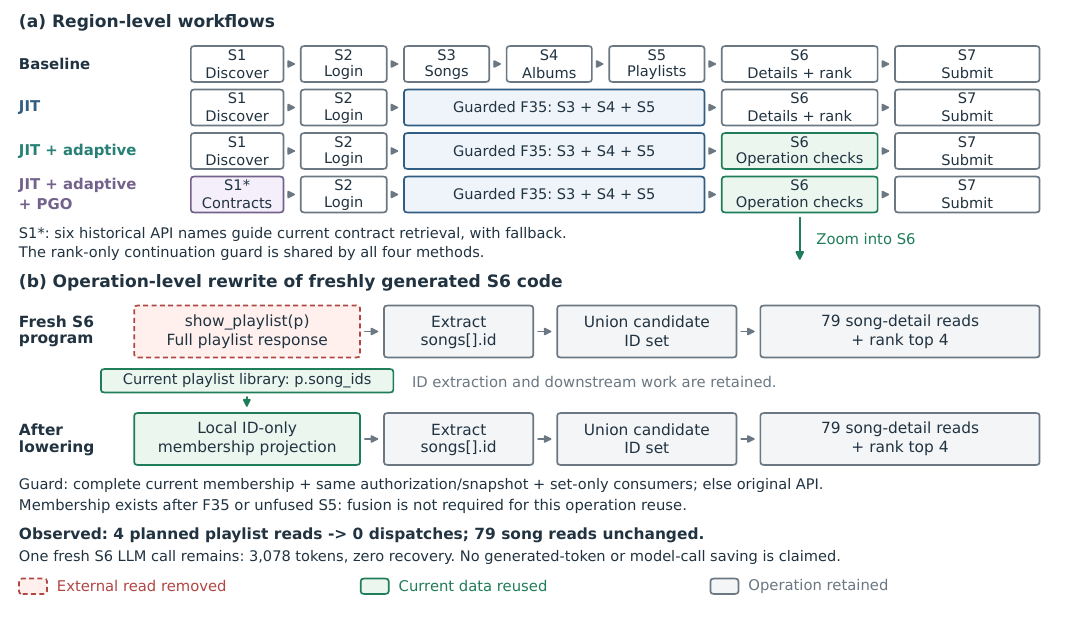}
  \caption{Region and operation views of the measured workflows. The S6
  expansion shows a certified read replacement, retaining ID extraction
  and ranking. Its membership input is also available after unfused S5;
  this operation rewrite does not require fusion.}
  \label{fig:motivation-information-workflows}
\end{figure*}

Consider an AppWorld task~\cite{appworld} that asks for the four most-played
R\&B song titles across a user's Spotify song, album, and playlist libraries.
Its seven-stage plan discovers APIs (S1), authenticates (S2), collects the
three libraries (S3--S5), gathers song details and ranks candidates (S6), and
submits the answer (S7). Figure~\ref{fig:motivation-information-workflows}
shows the region boundaries and expands the operation changes inside S6.

\paragraph{JIT: reduce generation boundaries visible in the plan.}
S3--S5 share authentication and repeat task and interface context when
generated separately. A region pass groups their duties into F35, guarded
by current contracts and authentication. DeepSeek generates their loops
together, retaining complete pagination and all three outputs. Collection
requires one model call instead of three: median cost falls from 7,015.5
to 3,588 tokens and from 3.12 to 1.50 seconds. Whole-task medians fall from
21,071 to 17,419 tokens and from 10.48 to 8.65 seconds. The opportunity is
fewer generation boundaries with the same required collection work.

\paragraph{Adaptive: check what the model actually proposes to execute.}
A compact plan does not ensure efficient generated actions. Some S6
programs fetch playlist details only to extract IDs already present in the
completed library. Figure~\ref{fig:motivation-information-workflows}(b)
expands one observed program: a local ID-only projection replaces four
\texttt{show\_playlist} calls. ID extraction and set accumulation remain;
the 79 song-detail reads and ranking still execute. Before lowering, the
operation pass certifies set-only use, complete membership, matching
contracts, and current data and authorization scope. Unknown fields,
order-sensitive consumers, or invalid state retain the original read.
A separate rule reuses identical GET results under the same scope.
Both rules use current evidence, not historical code or values.

Across the adaptive-enabled configurations, three certified rewrites remove
12 playlist reads; one repair reuses four equivalent reads. Post-generation
lowering cannot save tokens already spent on that code: JIT and
JIT+adaptive each use 11 S6 model calls over ten trials. Their median total
difference is only 83.5 tokens, and paired median token and latency intervals
include zero. These results demonstrate guarded read elimination, but not
an independent net token or latency advantage from adaptive optimization.

\begin{table*}[!tbp]
\centering
\setlength{\tabcolsep}{6pt}
\begin{tabular}{@{}lrrrrr@{}}
\toprule
Method & Tokens & LLM calls & Executor (s) & Recovery $\Sigma$ & Success \\
\midrule
Baseline & 21,071 & 8.5 & 10.48 & 27,415 & 10/10 \\
JIT & 17,419 & 6 & 8.65 & 31,822 & 10/10 \\
JIT + adaptive & 17,335.5 & 6 & 8.30 & 9,760 & 10/10 \\
JIT + adaptive + PGO & 12,920 & 5 & 8.34 & 11,812 & 10/10 \\
\bottomrule
\end{tabular}
\caption{Ten attempts per method on one frozen backend with fresh DeepSeek
code. Tokens, calls, and time are successful medians including all failed
calls and repairs. Recovery $\Sigma$ sums its included tokens over ten
attempts. Baseline is the matched unfused control. Common plan acquisition
(1,219 tokens) and independent profile acquisition (24,815 tokens) are separate.}
\label{tab:motivation-whole-task}
\end{table*}

\begin{figure*}[!tbp]
  \centering
  \includegraphics[width=\linewidth,height=0.75\textheight,keepaspectratio]{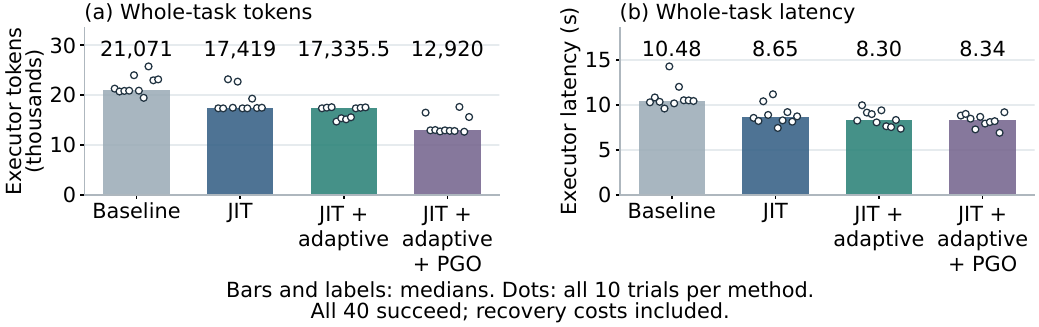}
  \caption{All trials from Table~\ref{tab:motivation-whole-task}; points show
  attempts and bars show medians. All succeed, with recovery included.
  PGO minimizes median tokens, but latency does not improve monotonically.}
  \label{fig:motivation-information-tokens}
\end{figure*}

\paragraph{PGO: transfer useful choices to the next execution.}
S1 must choose useful APIs again on a new run. Separate JIT+adaptive traces
identify six interfaces for login, library collection, membership, and song
details. PGO supplies their names as candidates; the new run fetches current
contracts, validates coverage, and falls back to discovery when needed.
Cold discovery also permits direct documentation requests. The profile
transfers interface choices, excluding old schemas, task values, and code.

S1 falls from a median 6,771 to 2,203.5 tokens and from two model calls to
one. The full configuration uses 12,920 tokens, 25.5\% fewer than
JIT+adaptive and 38.7\% fewer than baseline. Its median latency is
8.34 seconds, close to JIT+adaptive's 8.30 seconds. This implementation
transfers API choices, not a learned profitability model.

\paragraph{What the example motivates.}
The three information sources expose region grouping, concrete operation
reuse, and recurring interface choices. AgentPRO coordinates these through
a shared hierarchical IR: JIT before execution, adaptive checks during
execution, and PGO before a subsequent run. New facts and rewrites can
change dependencies and input liveness. This motivates revisiting affected
passes and analyses under a cost budget. The measured benefits are
conditional: fusion and discovery reduce model cost here, while adaptive
demonstrates operation reuse without established additional model savings.

\paragraph{Measurement scope.}
All methods share document forwarding, context projection, verified
closure, schema and authentication checks, recovery, and the preplanned
rank-only guard; that guard is not an adaptive contribution. Blocks are
randomized, with no replacement attempts or code changes. This single-task
example is separate from the 115-task evaluation in
Section~\ref{sec:experimental-results}.

\section{Approach Overview}
\label{sec:approach-overview}

\begin{figure*}[!tbp]
  \centering
  \includegraphics[width=\textwidth,height=0.75\textheight,keepaspectratio]{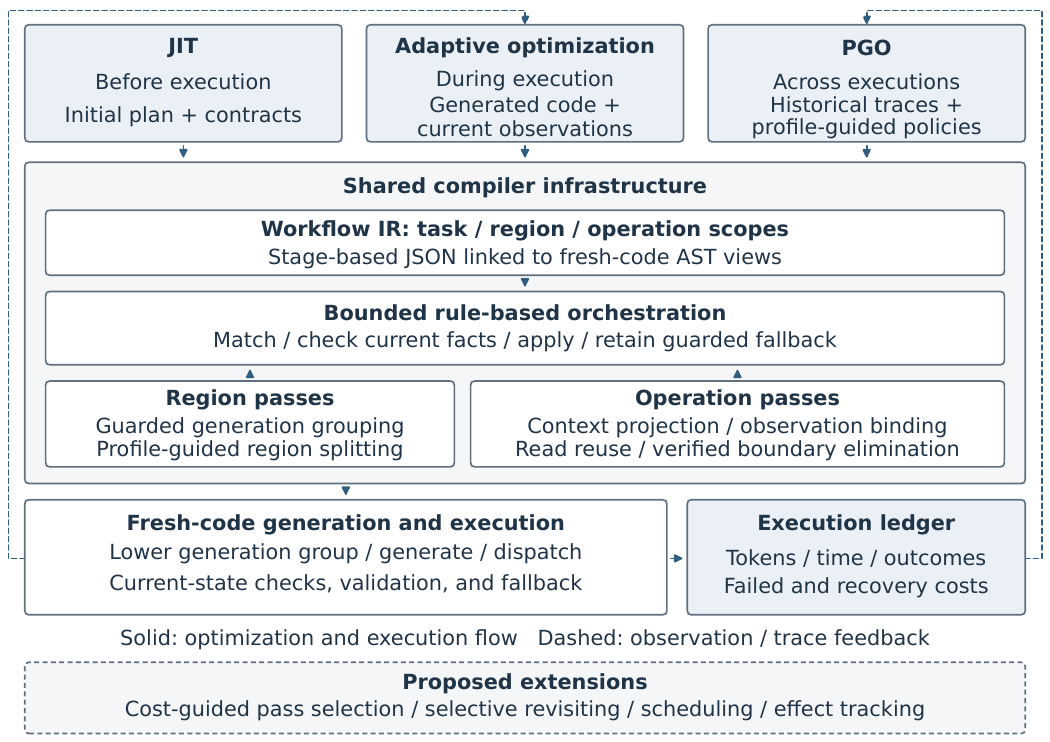}
  \caption{AgentPRO's shared optimization path. Independently selectable JIT,
  adaptive optimization, and PGO use initial plans, generated code and current
  observations, and historical traces to drive the same implemented pass library.
  Bounded rule-based orchestration checks applicability and retains guarded
  fallbacks. A history-guided policy can take effect during execution after
  current-state checks. The backend generates fresh code and records execution
  and recovery costs. Solid arrows show optimization and execution flow; dashed
  arrows show current observations and recorded traces feeding later decisions.
  The bottom panel separates proposed cost-guided pass selection, selective pass
  revisiting, and general scheduling and effect tracking from the implemented path.}
  \label{fig:agentpro-framework}
\end{figure*}

AgentPRO keeps an agent workflow available for optimization as its plan becomes
generated code, execution reveals current state, and completed runs provide
experience. Its objective is to reduce tokens, latency, or a specified trade-off
while preserving required work and external effects. Figure~\ref{fig:agentpro-framework}
shows the design and its implemented subset. JIT, adaptive optimization, and
PGO provide three information entry points to a shared IR and region and operation pass libraries.
They differ in the evidence available and the unfinished work they can change.

\paragraph{Represent the work and its evidence.}
The frontend builds a partial hierarchy of task, regions, and operations from
the plan and available tool contracts. The proposed IR records inputs, outputs,
effects, completion conditions, and execution-data and generation-observation
dependencies. Unresolved requirements remain explicit. Region containment,
joint code generation, and concurrent execution are separate decisions: fusion
must preserve internal operations and required observations. Facts retain their
provenance and validity conditions so that a historical choice cannot be
mistaken for a current value.

\paragraph{Optimize before execution with JIT.}
JIT uses plan structure and available contracts to construct an initial guarded
workflow. Region passes group compatible generation duties; operation passes
can simplify work supported by existing evidence. In the library example, JIT
prepares joint generation for S3--S5, enabled only after the required interfaces
and authentication are ready. Missing observations retain separate work or
guarded alternatives. JIT does not assume future collection contents or success.

\paragraph{Check concrete operations with adaptive optimization.}
Fresh generation exposes operations that the plan did not determine. Before
execution or tool dispatch, adaptive checks supported patterns against valid
current state. The shared operation passes bind observations and reuse eligible
reads. In the library example, the prototype can replace playlist reads used
solely for an unordered ID set with a local membership projection, and reuse eligible
identical GET results. Guards preserve the original operation when the evidence
or consumer is incompatible. ID extraction, song-detail acquisition, and ranking
remain. The broader design can revisit affected unfinished regions at safe
points while retaining valid progress and respecting committed effects.

\paragraph{Carry useful choices forward with PGO.}
Completed execution ledgers guide the selection and configuration of shared
passes for later runs, including observation binding and region splitting.
In the library example, the prototype extracts six useful API names from prior
JIT+adaptive traces and specializes the next discovery objective; execution
still fetches current contracts, validates coverage, and falls back to open
discovery. History-guided policies apply only when their current-state checks
hold, including when the required observations arrive during execution. The
proposed profile format additionally conditions cost, success, and recovery estimates on workflow,
model, and interface versions. Profiles guide decisions; each run acquires fresh
task data and generates fresh code. Profile acquisition costs are recorded
separately.

\paragraph{Coordinate and revisit passes.}
The proposed pass catalog includes grouping, splitting, and scheduling at
region scope, and read reuse, dead-work removal, and context projection at
operation scope. The proposed manager first
checks a pass's requirements, then estimates its total cost, including
optimization, regeneration, and recovery. A rewrite or new fact returns a change
set: affected analyses become invalid, and dependent passes re-enter a bounded
worklist. An earlier failed match can therefore become applicable. Reconsideration
stops when no justified gain remains or the budget is exhausted.

\paragraph{Lower, execute, and account.}
The backend lowers selected groups into prompts and contracts, inspects
supported operations in fresh generated code, and records tool observations,
outcomes, and every incurred model call. The measured prototype uses a shared
stage-based JSON IR and fresh-code AST views, with bounded rule-based
orchestration. Implemented passes include guarded grouping, profile-guided
splitting, context projection, observation binding, guarded read reuse, and
verified boundary elimination; the library example also specializes discovery.
General cost-guided selection, selective pass revisiting, arbitrary effect-safe
recompilation, and cross-task transfer remain proposed or unevaluated.

\section{Approach Details}
\label{sec:approach-details}

We describe AgentPRO using the library task in
Figure~\ref{fig:motivation-information-workflows}. The agent must collect three libraries,
obtain song details, and return the four most-played R\&B songs. AgentPRO
separates this process into three components: a frontend that represents the
work and its evidence, an optimizer that changes regions and operations, and
an execution backend that generates fresh code and returns observations.
JIT, adaptive optimization, and PGO invoke these components with different
information. We first define the design, then identify the implemented path
in each component. Table~\ref{tab:prototype-scope} summarizes the boundary
between the current prototype and proposed extensions.

\subsection{Constructing the Workflow IR}
\label{sec:input-extraction}
\label{sec:trace-lifting}
\label{sec:hierarchical-ir}

\paragraph{Inputs and hierarchy.}
The frontend accepts a task, its generated plan, available tool contracts, and
optional historical profiles. A task can start without an execution trace.
The proposed IR has three scopes: a task contains regions, and each region
contains operations or nested regions. A region represents a duty such as
collecting a library; an operation represents a model request, tool access,
local computation, validation, or completion. Regions declare required inputs,
produced values, effects, and completion conditions. Operations record their
identity, arguments, results, parent region, and execution state. An unresolved
plan step retains an objective and requirements until generation supplies its
concrete body.

Generation groups specify which duties are presented in one code-generation
request. They are separate from region containment and execution scheduling.
For example, S3--S5 can share generation while their collection loops execute
in their original order. A region can also contain several model turns when
later generation requires an intermediate observation. Fusion therefore does
not imply one model call under every execution path or concurrent tool access.

\paragraph{Dependencies and evidence.}
The IR distinguishes three requirements. An execution-data dependency supplies
a value needed to run an operation. A generation-observation dependency
supplies evidence needed to decide what code to generate. Control and effect
constraints preserve branching, required ordering, and external actions.
A collection loop can be generated with a token parameter before the token's
value is known; a choice that depends on an unseen response must retain that
observation boundary. Unknown effects or ambiguous uses constrain rewrites
conservatively.

Each fact records its source and validity conditions. Current tool results,
interface contracts, and historical measurements have different authority:
a previous successful read cannot certify a current value. The proposed runtime
tracks fact versions, live values, and operation states, and records committed
external effects separately. These records let an optimizer identify eligible
unfinished work without replaying completed actions.

\paragraph{Implemented representation.}
The prototype uses a bounded JSON plan IR with stable stage identifiers,
objectives, source regions, interfaces, guards, and unresolved requirements.
Fresh Python code supplies an AST view for supported operation checks.
It does not implement automatic extraction of a complete dependency graph or
a general versioned value store and effect journal.

\subsection{Region and Operation Passes}
\label{sec:analyses-contracts}

AgentPRO optimizes workflows at two complementary scopes.
Region passes determine how work is grouped and scheduled;
operation passes reduce the work and information processed within
that structure. Table~\ref{tab:pass-library} summarizes the shared
pass library, including proposed extensions; Table~\ref{tab:prototype-scope}
identifies the implemented subset.

\paragraph{Region passes.}
Generation fusion combines compatible pending regions into one
generation group. It preserves their IR operations, required
observations, and completion conditions; removing redundant
operations remains the responsibility of operation passes.
Fusion can reduce generation boundaries and repeated prompt
context. Interface narrowing removes exports that have no live
consumers. Region splitting divides pending groups to bound
generation and repair scope. Parallel scheduling overlaps
independent work while preserving generation dependencies,
required observations, and effect ordering.

\paragraph{Operation passes.}
Operation passes rewrite individual tool, prompt, and controller
actions. Context projection removes irrelevant static prompt
examples while preserving protocol information and live inputs.
Observation binding connects pending inputs to facts already
established in the current execution. Equivalent-read elimination
replaces a repeated read with a valid current-run result.
Dead-operation elimination removes unused pure operations, while
contract specialization and result-field projection retain only
the interface information and fields required by consumers.
Verified boundary elimination removes an acknowledgment after
the goal's postcondition has been verified. Retry-checkpoint
reuse preserves a valid successful prefix without replaying
committed effects.

These transformations require more than syntactic similarity.
Read reuse checks API identity, arguments, authorization, state,
and freshness. Deletion and projection require complete consumer
information, including later prompts, verification, and exported
state. A successful tool invocation alone does not justify
closing a region: its result must satisfy the region's completion
condition.

\paragraph{Shared applicability and composition.}
Each pass specifies its match conditions, required facts, legality
checks, rewrite, and invalidated analyses. JIT, adaptive optimization,
and PGO invoke eligible passes from this library using planning
information, current observations, and historical profiles,
respectively. Profiles identify and prioritize candidates but
do not replace current validity checks. A rewrite can change the
applicability or expected cost of another pass, so the proposed selective manager
reconsiders affected passes when relevant facts or IR properties
change. When a legality condition cannot be established, the
original pending work is retained. Profitability is evaluated
separately, accounting for optimization, generation, execution,
and recovery costs.

\begin{table*}[!tbp]
\centering

\setlength{\tabcolsep}{4pt}
\renewcommand{\arraystretch}{1.0}
\begin{tabularx}{\textwidth}{@{}l>{\raggedright\arraybackslash}p{0.25\textwidth}>{\raggedright\arraybackslash}X@{}}
\toprule
ID & Pass & Transformation and applicability \\
\midrule
\multicolumn{3}{@{}l}{\textit{Region passes}} \\
R1 & Generation fusion
   & Fuse compatible pending generation regions while preserving
     required observations, operations, and completion conditions. \\
R2 & Interface narrowing
   & Remove unused region exports when all consumers,
     including prompts and verification, are known. \\
R3 & Region splitting
   & Split pending generation groups to bound generation
     and repair scope without repeating completed work. \\
R4 & Parallel scheduling
   & Overlap independent work while preserving generation
     dependencies, required observations, and effect ordering. \\
\midrule
\multicolumn{3}{@{}l}{\textit{Operation passes}} \\
O1 & Context projection
   & Remove irrelevant static prompt examples while retaining
     protocol demonstrations and the exact live input. \\
O2 & Observation binding
   & Bind pending observation inputs to validated current-run
     facts instead of requesting them again. \\
O3 & Equivalent-read elimination
   & Reuse an equivalent read result when API identity, arguments,
     authorization, state, and freshness requirements match. \\
O4 & Dead-operation elimination
   & Delete unused pure operations when consumers and effects
     are fully known. \\
O5 & Contract specialization
   & Replace documentation preparation with a compact current
     API contract when all required fields are available. \\
O6 & Result-field projection
   & Remove unused result fields while retaining fields required
     for identity, pagination, and verification. \\
O7 & Verified boundary elimination
   & Remove an acknowledgment boundary after fresh execution
     satisfies the region's verified postcondition. \\
O8 & Retry-checkpoint reuse
   & Reuse a successful same-run prefix when its outputs,
     authorization, and state remain valid. \\
\bottomrule
\end{tabularx}
\caption{AgentPRO's shared region and operation pass library.
Each pass applies when its required dependency, effect,
and validity conditions hold. Table~\ref{tab:prototype-scope} distinguishes
implemented mechanisms from proposed extensions.}
\label{tab:pass-library}
\end{table*}
\subsection{JIT: Building the Initial Workflow}
\label{sec:jit-optimization}

JIT runs after planning. It lifts the plan into the partial IR and considers
transformations supported by the plan and available contracts. Its output is
an initial workflow with generation groups, required observations, and guards.
Missing runtime facts remain explicit. Compatible historical profiles can rank
candidates, but cannot supply current task values.

For the library task, the implemented compiler reads the original seven-step
plan and checks the collection duties' authentication, pagination, and output
requirements. It prepares F35 to collect the song, album, and playlist
libraries together. A readiness guard checks current interfaces and
authentication before joint generation; otherwise, execution retains the
separate stages. F35 preserves the listed collection order, complete paging,
and any necessary documentation or observation steps. On the measured normal
collection path, three generation requests become one.

The benefit comes from avoiding repeated generation context and turns. JIT
cannot certify the contents of libraries that have not been read, nor the
operations that a future model response will contain. Selecting a prepared
guarded alternative when its conditions become true executes the JIT workflow;
it is distinct from rewriting newly generated operations.

\subsection{Adaptive: Checking Work Before Dispatch}
\label{sec:adaptive-optimization}

Adaptive optimization receives information unavailable when the plan was
compiled: the actual generated program and validated current observations.
The prototype applies eligible shared passes using this evidence, including
observation binding and guarded read reuse. In the library example, it runs
conservative AST checks after fresh S6 generation and before execution, and
checks eligible reads at dispatch. It does not ask
another model to judge whether arbitrary actions are necessary.

\paragraph{A concrete operation rewrite.}
In the observed S6 program, the model reads each playlist to extract song IDs
and inserts those IDs into an unordered candidate set. The current playlist
library already supplies membership. The supported rewrite changes the source
of that ID view:
\begin{quote}

\textbf{Before:} \texttt{read\_playlist(p)}\\*
\hspace*{1em}$\rightarrow$
\texttt{extract\_ids} $\rightarrow$ \texttt{set\_union}.\\
\textbf{After:} \texttt{membership\_id\_view(p)}\\*
\hspace*{1em}$\rightarrow$
\texttt{extract\_ids} $\rightarrow$ \texttt{set\_union}.
\end{quote}
The replacement constructs only the response view used by the ID consumer.
It retains the original API call as the missing-key fallback. ID extraction,
set union, song-detail acquisition, and ranking remain; in the illustrated
execution, the 79 song-detail reads still occur.

The AST check requires that the response be used only through membership IDs
accumulated into a set. Unsupported aliases, exported response objects,
mutations, ordered consumers, or unrecognized uses cause abstention. Current
membership, authorization, and data scope must also be valid. This rule relies
on a pinned backend-specific interface/source contract and a Boolean
referential-integrity check on current membership. The check returns no task
IDs or answers, and does not replace consumer or authorization checks.
The resulting certificate justifies this restricted observation, not
full-response equivalence for arbitrary API clients.

\paragraph{Reuse and reconsideration.}
Eligible identical GET results can also be reused within a valid current scope.
The prototype keys results by operation, arguments, method, and scope, caches
a result only when the scope remains unchanged across the read, and returns
a copy. Unsupported or invalid entries execute the original request.
The same membership evidence can exist after an unfused collector, so fusion
is not a prerequisite for this rewrite.

The general adaptive design additionally invokes the pass manager at safe
points when facts, sizes, or dependencies change. It can revise affected
unfinished regions while retaining valid progress. In-flight operations and
committed effects constrain such changes. The prototype's bounded checks do
not implement this general recompilation mechanism. Because code generation
has already occurred, a post-generation rewrite cannot reclaim its tokens.
Its net token or latency benefit must include checking overhead and any later
execution or model work it avoids.

\subsection{PGO: Specializing Subsequent Workflows}
\label{sec:pgo-optimization}

PGO uses completed traces to select and configure shared region and operation
passes for later executions. The evaluated mechanisms include observation
binding and region splitting, as well as context, read-reuse, and closure
policies. In the proposed profile format, a profile associates workflow features
and candidate transformations with model, interface, and backend versions;
observed tokens, latency, success,
and recovery characterize their cost. Failed attempts and guard failures remain
in the profile. Sparse or incompatible evidence reduces confidence rather than
relaxing legality checks. These profiles can guide both initial grouping and
later adaptive decisions.

The library-task PGO example specializes API discovery. Two independent prior
JIT+adaptive runs identify six useful API names for the library task. The next
run receives these names as candidates, fetches their current contracts, checks
coverage, and falls back to open discovery when needed. Thus history changes
the discovery objective before new execution begins. It imports neither an old
schema nor historical code or task values. The model still generates fresh
code and collects current data.

A history-guided policy may be selected before execution and take effect only
after current observations establish its validity. For example, PGO can select
an observation-binding policy whose inputs are bound after the current run
produces the required facts. Adaptive optimization instead makes a new decision
from current execution evidence. Both use eligible shared passes and current
validity checks; neither imports historical answers or certifies stale values.

Profile construction precedes the evaluated runs; a run cannot benefit
retroactively from its own trace. Profile acquisition and policy-building costs
are reported separately from executor cost, with a reuse horizon required for
amortized claims. The library experiment tests API-choice transfer on one task;
the broader evaluation exercises history-guided shared passes across the
baseline-qualified cohort. Learned transformation costs and validated transfer
to tasks absent from the profile remain proposed.

\subsection{Selecting and Revisiting Passes}
\label{sec:pass-selection}
\label{sec:pass-revisiting}

The proposed manager separates applicability from profitability. Rules first
check a candidate's dependencies, effects, and validity requirements. A cost
model then estimates whether changing the remaining workflow is worthwhile.
Token mode minimizes expected tokens; latency mode minimizes completion time
under a token budget. A joint mode uses fixed reference values and weights:
\begin{equation}
 \widehat J(\mathcal I)=
 w_T\frac{\widehat T(\mathcal I)}{T_{\mathrm{ref}}}
 +w_L\frac{\widehat L(\mathcal I)}{L_{\mathrm{ref}}}.
\end{equation}
Estimates include optimization, generation, tools, validation, and recovery.
Latency follows the critical path rather than summing overlapping durations.
A candidate must exceed an uncertainty margin: fewer model calls or smaller
prompts can still increase generated code and repair cost. Without compatible
history, conservative rules or abstention replace unsupported precise estimates.

A worklist entry identifies a pass, its target, and relevant fact versions.
When a rewrite or runtime observation changes those facts, the manager
invalidates affected analyses and re-enqueues dependent candidates through
containment and dependency links~\cite{llvm-npm,mlir-rewriting}. For example,
removing redundant documentation preparation can remove an observation
requirement that previously blocked joint generation. Fusion is then checked
again on the affected collection regions. A notification prompts reanalysis;
it does not itself authorize a rewrite. Candidate and time budgets, repeated
IR-state checks, and a minimum expected gain bound reconsideration.

This selective manager is proposed. The prototype applies bounded rules rather
than a learned cost policy or a general invalidation worklist. The earlier O/R
audit separately showed that removing preparation before fusion yields a
two-slot collection schedule, whereas the reverse order leaves three slots;
a static fusion revisit reaches two. These IR counts demonstrate applicability
changes, not measured performance of the proposed revisiting manager.

\subsection{Code Generation, Feedback, and Prototype Scope}
\label{sec:lowering-execution}
\label{sec:outputs-feedback}

The backend lowers each selected generation group into a prompt containing its
objective, current interfaces and inputs, required outputs, and completion
checks. The model generates fresh code. Supported operation checks may rewrite
that new code before dispatch; they do not replace generation with a stored
implementation or a workflow interpreter. Tools supply actual observations,
which are forwarded to their consumers and to the next optimization decision.

The execution ledger records all incurred model calls, tool events, timing,
cache hits, outcomes, and recovery. Failed or rejected calls and successful
repair both contribute to total cost. In the illustrative library-task experiment,
common setup, context projection, completion handling, and the rank-only
continuation guard are shared by the
four configurations. Structural adherence and task success are
reported separately; neither proves equivalence of arbitrary generated code.

Table~\ref{tab:prototype-scope} identifies the implemented mechanisms. The
illustrative matched library experiment demonstrates gains for the full
configuration and concrete read eliminations; it does not isolate an independent
net adaptive token or latency benefit. The broader evaluation reports realized
shared-pass effects and distinguishes historical comparisons from matched
JP/AP/JAP runs. Neither study establishes the performance of a general
cost-guided pass manager or transfer to tasks absent from the profile.

\begin{table*}[!tbp]
\centering

\setlength{\tabcolsep}{4pt}
\begin{tabularx}{\textwidth}{@{}>{\raggedright\arraybackslash}p{0.14\textwidth}>{\raggedright\arraybackslash}X>{\raggedright\arraybackslash}X@{}}
\toprule
\textbf{Component} & \textbf{Current prototype} & \textbf{Proposed extension} \\
\midrule
IR & Stage-based JSON IR linked to fresh-code AST views, with unresolved requirements retained & General dependency analysis, versioned values, and an effect journal \\
Region passes & Guarded generation grouping and profile-guided region splitting & General scheduling and overlap under complex dependency and effect constraints \\
Operation passes & Context projection, observation binding, guarded read reuse, and verified boundary elimination & General dead-operation elimination and broader consumer-specific rewrites \\
JIT & Uses the initial plan and available contracts to apply eligible shared passes and prepare guarded alternatives & Cost-guided selection under incomplete execution information \\
Adaptive optimization & Uses generated code and current observations to apply eligible shared passes, including guarded rewrites before dispatch & Selective reoptimization of unfinished regions while preserving committed effects \\
PGO & Uses historical traces to select and configure shared passes, with current-state checks governing application & Learned cost and reliability estimates and validated transfer across tasks \\
Pass manager & Bounded rule-based orchestration; pass-order interactions and revisiting examined through a static audit & Cost-guided selection, incremental analysis invalidation, and selective pass revisiting under a budget \\
\bottomrule
\end{tabularx}
\caption{Implemented mechanisms and proposed extensions. JIT, adaptive optimization, and PGO use shared region and operation passes, guided by initial plans, current observations, and historical traces, respectively. Listed mechanisms are representative; each application remains subject to the pass's dependency, effect, and validity conditions.}
\label{tab:prototype-scope}
\end{table*}

\section{Evaluation Setup}
\label{sec:evaluation-setup}

\paragraph{LLM.}
We use DeepSeek~\cite{deepseek}, with the provider-reported model identifier \texttt{deepseek-flash}, for planning and code generation. In the cross-task evaluation, we set the temperature to 0, disable thinking, and limit each response to 400 output tokens. Configurations within each matched experiment use identical model settings.

\paragraph{Dataset.}
We evaluate on AppWorld~\cite{appworld}, which provides stateful application APIs and task-specific correctness evaluators. Our workload contains the 115 tasks that succeeded in an initial baseline execution of all 168 tasks in the \texttt{test\_normal} split, covering 49 scenarios. Reported success rates therefore describe this baseline-qualified cohort.

\paragraph{Agent framework.}
The baseline follows AppWorld's official Plan-and-Execute workflow~\cite{appworld-plan-execute}: a planner produces a task plan, and an executor generates Python code to interact with application APIs and observes the results. AgentPRO optimizes the corresponding workflow IR using its shared region and operation pass library. Compared configurations use stored original plans and fresh execution code. The cross-task execution limits are 100 model calls, 500 environment calls, and 600 seconds per execution.

\paragraph{Comparisons.}
The overall comparison uses one stored attempt per task for the baseline, Automatic AgentPRO's heuristic controller, and the always-on policy. The lifecycle studies use three repetitions per task and configuration. JP, AP, and JAP share one matched batch with counterbalanced order, backend, plans, and profile; P uses that backend in an earlier batch, while J, A, and JA retain their original results. The baseline and original lifecycle study use earlier backend versions, so cross-batch comparisons are historical. The learned selector is evaluated offline on the original seven-configuration matched outcomes, without replacing them with the later PGO runs. Its fixed JAP comparator is a lifecycle configuration under the automatic manager, distinct from the overall comparison's always-on policy.

\paragraph{Metrics.}
We measure official task success rate, token consumption, component latency, and the numbers of model and environment calls. Token consumption includes input and output tokens, including cached input. We report all-attempt mean costs and successful-run medians unless explicitly labeled otherwise, retaining failed executions and all recovery work in the primary totals. Component time includes model requests, environment execution, and online optimization; embedded optimizer time is counted once. Physical process wall time additionally includes setup and grading and is reported separately. We record optimizer token usage and exclusive wall and CPU time for JIT, adaptive optimization, and PGO. Recovery totals distinguish failed/rejected calls from non-failing repair or fallback continuations; a non-failing call need not complete its recovery episode. Stored planning costs are attributed consistently across configurations; physical new executor spending and historical profile acquisition are reported separately.

\section{Experimental Results}
\label{sec:experimental-results}

\subsection{Overall Performance}
\label{sec:overall-performance}

Automatic AgentPRO reduces mean token consumption by 41.85\% and mean component latency by 6.61\% relative to the archived baseline on the 115-task cohort (Table~\ref{tab:overall-performance}). It completes 110 tasks (95.7\%), while the always-on policy completes 107 (93.0\%). The baseline completes all 115 tasks by construction: these tasks were selected from its successful executions. Each method has one recorded attempt per task. The automatic policy in this comparison is the heuristic controller; the learned selector is evaluated separately below.

\begin{table}[!tbp]
\centering
\setlength{\tabcolsep}{2pt}
\begin{tabularx}{\columnwidth}{@{}>{\raggedright\arraybackslash}Xrrr@{}}
\toprule
Method & \shortstack{Stored\\baseline} & \shortstack{Automatic\\AgentPRO} & Always-on \\
\midrule
Success & 115/115 & 110/115 & 107/115 \\
\shortstack[l]{Mean tokens\\($10^3$)} & 180.38 & 104.89 & 108.83 \\
\shortstack[l]{Token saving\\(\%)} & -- & 41.85 & 39.67 \\
Mean time (s) & 35.28 & 32.95 & 33.88 \\
\shortstack[l]{Time reduction\\(\%)} & -- & 6.61 & 3.97 \\
\shortstack[l]{Median tokens\\($10^3$)} & 163.07 & 89.89 & 91.83 \\
\shortstack[l]{Median time\\(s)} & 31.35 & 29.23 & 30.15 \\
\bottomrule
\end{tabularx}
\caption{Historical comparison on all 115 baseline-qualified tasks. Percentages use all-attempt means relative to the stored baseline; medians use successful attempts. Costs retain planning, failed calls and recovery. The earlier baseline protocol prevents an identical-backend causal interpretation.}
\label{tab:overall-performance}
\end{table}

\begin{figure*}[!tbp]
\centering
\includegraphics[width=\linewidth,height=0.75\textheight,keepaspectratio]{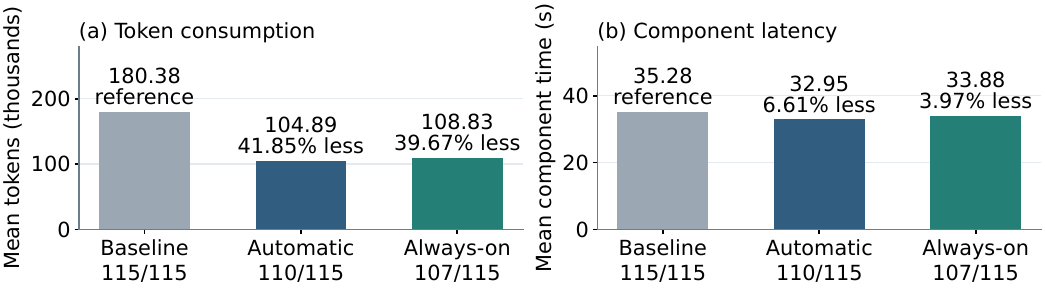}
\caption{Overall token and latency costs on the 115-task historical cohort. Annotations give reductions relative to the archived baseline. Counts under the methods show official successes. All registered failure and recovery costs remain included.}
\label{fig:overall-performance}
\end{figure*}

Always-on reduces mean tokens by 39.67\% and mean component latency by 3.97\% relative to the same archived baseline. These cost reductions accompany five and eight failed tasks for automatic and always-on, respectively; they do not establish success-rate preservation. Failed tasks remain in the headline means. Aggregate recovery costs are 2.431 million tokens for the baseline, 1.780 million for automatic, and 1.882 million for always-on. Automatic's recovery cost comprises 0.336 million tokens in failed or rejected calls and 1.444 million in non-failing recovery calls; both are already included in its total.

On the 107 tasks completed by all three methods, automatic reduces aggregate tokens by 44.0\% and component time by 10.2\%. This common-success comparison is a secondary, outcome-conditioned analysis. Of its 8.111-million-token reduction, 8.109 million tokens are input tokens. The trace audit records context projection on all 107 tasks, 144 eliminated authentication acknowledgments, equivalent-read reuse on 78 tasks, and fused dispatches on 19 tasks. These observations identify concrete changes in prompts and execution, but do not isolate each pass's net contribution.

The automatic and always-on arms have identical initial request payloads and dispatched generation groups on all 107 common-success tasks. Their observed performance gap therefore does not demonstrate a benefit from policy selection. Neither arm applies a PGO rewrite because no compatible profile is available in this comparison. The lifecycle comparison below reports all seven capability configurations and incorporates the completed shared-library PGO evaluations.

\subsection{Performance of JIT, Adaptive Optimization, and PGO}
\label{sec:lifecycle-results}

AgentPRO reduces mean token consumption by 38.64--42.12\% across the seven configurations in Table~\ref{tab:lifecycle-results}. J, A, and P denote JIT, adaptive optimization, and PGO. JAP achieves the lowest mean token cost, 104,403 tokens per attempt, and completes 331/345 executions. PGO alone achieves 107,670 tokens and the same success count, with the lowest recorded component latency of 32.09 seconds. Thus, the best token configuration and the best latency configuration differ.

The table uses the latest completed P, JP, AP, and JAP evaluations with the shared pass library. JP/AP/JAP form 345 matched three-arm blocks with counterbalanced order, identical stored plans, and a frozen profile; P uses the same backend in an earlier batch. J, A, and JA retain their original recorded results. All methods cover the same 115 baseline-qualified tasks. Baseline-relative percentages are historical comparisons; the matched analysis below uses JP/AP/JAP. The profile contains prior same-task history for 107 tasks, while eight lack that history.

\begin{table*}[!tbp]
\centering

\setlength{\tabcolsep}{3.5pt}
\begin{tabular}{@{}lrrrrrrr@{}}
\toprule
Configuration & Success & \shortstack{Mean\\tokens\\($10^3$)} & \shortstack{Token\\saving\\(\%)} & \shortstack{Mean\\time\\(s)} & \shortstack{Time\\reduction\\(\%)} & \shortstack{Optimizer\\(s/attempt)} & \shortstack{Recovery\\($10^3$/attempt)} \\
\midrule
Plan-and-Execute & 115/115 & 180.38 & -- & 35.28 & -- & -- & 21.138 \\
J & 321/345 & 105.42 & 41.56 & 32.45 & 8.04 & 0.189 & 18.705 \\
A & 317/345 & 107.94 & 40.16 & 33.49 & 5.07 & 1.091 & 18.436 \\
$P^{\dagger}$ & 331/345 & 107.67 & 40.31 & 32.09 & 9.05 & 0.263 & 23.003 \\
JA & 321/345 & 108.33 & 39.94 & 34.36 & 2.62 & 1.491 & 19.041 \\
$JP^{\dagger}$ & 328/345 & 105.90 & 41.29 & 32.26 & 8.57 & 0.275 & 24.792 \\
$AP^{\dagger}$ & 323/345 & 110.69 & 38.64 & 32.83 & 6.95 & 0.531 & 24.607 \\
$JAP^{\dagger}$ & 331/345 & 104.40 & 42.12 & 32.72 & 7.28 & 0.522 & 20.510 \\
\bottomrule
\end{tabular}
\caption{All-attempt performance relative to the stored Plan-and-Execute baseline. The baseline has one stored execution per task; optimized configurations have three. Recovery includes failed/rejected calls and non-failing repair or fallback continuations. Optimizer time is already included. $\dagger$ P/JP/AP/JAP use the latest backend/profile; J/A/JA retain the original results. JP/AP/JAP are matched within their new batch; P and baseline are historical references.}
\label{tab:lifecycle-results}
\end{table*}

\begin{figure*}[!tbp]
\centering
\includegraphics[width=\linewidth,height=0.75\textheight,keepaspectratio]{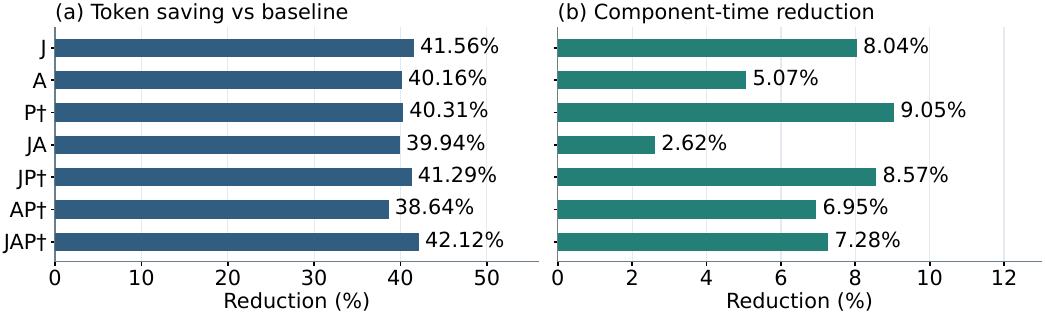}
\caption{Token and component-time reductions relative to the stored Plan-and-Execute baseline. All failed executions and recovery costs are included. $\dagger$ P, JP, AP, and JAP use the latest shared-library PGO evaluations; J, A, and JA retain the original results.}
\label{fig:lifecycle-reductions}
\end{figure*}

\paragraph{Combining mechanisms changes the cost--completion trade-off.}
Within the matched experiment, JAP consumes 1.41\% fewer tokens than JP and 5.68\% fewer than AP, while completing three and eight more executions, respectively. JP is the fastest of these three configurations: 32.26 seconds, compared with 32.72 for JAP and 32.83 for AP. Table~\ref{tab:matched-lifecycle-contrasts} reports scenario-cluster intervals, retaining sibling tasks and repeats together. The central result is a trade-off: JAP gives the best observed token--completion combination, while JP gives lower component time and optimization overhead.

\begin{table*}[!tbp]
\centering

\setlength{\tabcolsep}{3.5pt}
\begin{tabular}{@{}lrrr@{}}
\toprule
Comparison & \shortstack{Token reduction (\%)\\{[95\% interval]}} & \shortstack{Time reduction (\%)\\{[95\% interval]}} & \shortstack{Success difference (pp)\\{[95\% interval]}} \\
\midrule
JAP vs. JP & +1.41 [-1.88, +5.97] & -1.41 [-5.37, +2.34] & +0.87 [-0.54, +2.48] \\
JAP vs. AP & +5.68 [-0.22, +12.90] & +0.35 [-3.90, +4.61] & +2.32 [+0.00, +4.97] \\
JP vs. AP & +4.33 [+0.41, +8.80] & +1.74 [-1.66, +5.04] & +1.45 [-0.28, +3.13] \\
\bottomrule
\end{tabular}
\caption{Matched contrasts on all 345 blocks. Positive reductions favor the first configuration. Pointwise 95\% percentile intervals use 10,000 bootstrap samples of the 49 scenarios, retaining their tasks and repetitions.}
\label{tab:matched-lifecycle-contrasts}
\end{table*}

\paragraph{Online optimization is inexpensive.}
The optimizers make no additional LLM requests. Mean exclusive optimizer wall time is 0.275 seconds for JP, 0.531 for AP, and 0.522 for JAP, or 0.85\%, 1.62\%, and 1.60\% of component time. PGO alone requires 0.263 seconds. These costs are already included in the latency measurements. Resource savings therefore do not depend on ignoring the work performed by the optimizer.

\subsection{Where the Savings Come From}
\label{sec:corrected-pgo-results}

\begin{figure*}[!tbp]
\centering
\includegraphics[width=\linewidth,height=0.75\textheight,keepaspectratio]{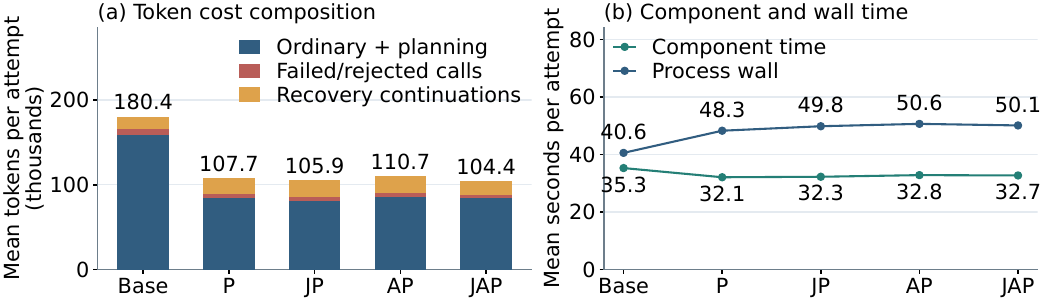}
\caption{Cost composition and timing for the latest PGO configurations. Token stacks partition each all-attempt mean into ordinary execution plus attributed planning, failed/rejected calls, and non-failing recovery continuations. Component time includes online optimization. Physical process wall time additionally includes setup and grading; the two time series are separate measurements, not additive components. Baseline and P are stored references; JP/AP/JAP are matched.}
\label{fig:corrected-pgo-results}
\end{figure*}

\paragraph{Broad savings and incremental savings have different sources.}
Relative to the stored baseline, JAP reduces mean cost by 75,975 tokens per attempt. Of this difference, 75,347 tokens (99.17\%) are outside recovery; recovery contributes the remaining 628 tokens. Figure~\ref{fig:corrected-pgo-results} shows the same broad pattern for the other PGO configurations: the large reduction is in ordinary execution, even when recovery costs increase. This locates the main opportunity in the repeated work of carrying context and generating actions, rather than only in repairing errors.

\paragraph{The hierarchy addresses opportunities at two scales.}
The latest JAP executions apply 9,743 context projections, reuse 942 reads, and remove 402 verified acknowledgments. They also apply 1,168 observation-binding rewrites, 39 generation-fusion rewrites, and 21 profitability splits. Operation passes act repeatedly on prompts, observations, and reads; region passes change the less frequent decisions about grouping or partitioning generation. Both scales are exercised within the same workflow representation.

The lifecycle determines which evidence drives these actions. In JAP, JIT supplies the 39 fusion rewrites; adaptive optimization supplies 803 observation-binding rewrites; and PGO supplies another 365 bindings and the 21 splits, alongside context, read-reuse, and closure policies. PGO can install a policy whose guarded effect occurs during execution. These counts describe realized changes, rather than candidate installations or an additive allocation of token savings. The important distinction is when useful evidence becomes available, not a separate pass library for each mechanism.

\paragraph{Recovery explains the smaller gaps between optimized configurations.}
Once the shared optimizations have reduced ordinary execution cost, recovery becomes decisive. JAP uses 1.477 million fewer recovery tokens than JP, offset by 0.961 million more ordinary-execution tokens, for a net reduction of 0.516 million tokens. Against AP, JAP saves 2.168 million tokens: 1.413 million from recovery and 0.754 million from ordinary execution. Recovery therefore accounts for 65.2\% of the JAP--AP gap. Recovery includes failed/rejected calls and non-failing continuations undertaken during repair or fallback. A non-failing call need not resolve its recovery episode; task completion is evaluated separately.

\begin{figure*}[!tbp]
\centering
\includegraphics[width=\linewidth,height=0.75\textheight,keepaspectratio]{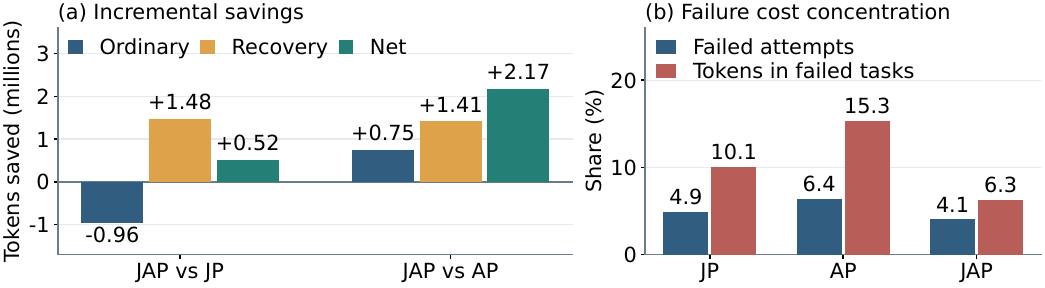}
\caption{Why workload means differ. Left: decomposition of JAP's token savings relative to JP and AP; negative ordinary-work savings indicate an increase. Identical planning attribution cancels. Right: the fraction of executions that fail and the fraction of total tokens spent on those failed executions. Every attempt remains included; whole failed-task expense overlaps the recovery categories.}
\label{fig:lifecycle-recovery-tail}
\end{figure*}

\paragraph{A small number of difficult paths can determine the mean.}
The all-attempt median token costs of JP, AP, and JAP are close: 91.32K, 90.80K, and 90.75K. Their failed executions consume 3.674M, 5.853M, and 2.259M tokens, respectively. AP's 6.4\% failed attempts account for 15.3\% of its total token cost. On the 316 blocks where all three configurations succeed, mean costs are 97.28K, 98.08K, and 98.30K: the ranking differs from the all-attempt mean. Thus, ordinary successful paths and expensive exceptional paths should be evaluated separately, with the latter retained in the primary objective.

The stored traces make this distinction concrete. In the third repetition of a corrected-rent-payment task (\texttt{9dabbc9\_3}), JAP uses 239.63K tokens and passes all nine tests, versus 388.70K for successful JP and 354.60K for failing AP. JAP binds newly available observations and sends the correction once; AP sends it twice. In a payment-funding task (\texttt{2c544f9\_2}), by contrast, all three configurations fail in the third repetition, but AP spends 1.127M tokens searching for funds while JAP stops after 140.90K. Across three repeats, this latter task contributes 0.649M tokens to JAP's 0.516M net advantage over JP, with no JIT/adaptive rewrite change in the representative run. The useful lesson is to optimize retained state, completion, and the length of unsuccessful search together: a shorter failed path and a successful repair are different outcomes.

\paragraph{Token reduction and latency reduction target different resources.}
In JAP, input tokens account for 98.12\% of fresh executor usage, and cached input accounts for 84.86\% of all executor tokens. JP and JAP make almost the same number of fresh requests, 9,735 and 9,743, and generate 663K and 666K output tokens. Reducing prompt volume can therefore leave the amount of generation and the number of round trips almost unchanged. The measured JAP--JP component-time difference is 0.455 seconds: 0.209 seconds of model time, $-0.002$ seconds of environment time, and 0.248 seconds of optimization outside environment execution. This explains why their token and latency rankings diverge without requiring a large optimizer cost.

Mean physical process wall time is 49.83, 50.65, and 50.10 seconds for JP, AP, and JAP, including setup and grading; P records 48.27 seconds and the historical baseline 40.57 seconds. These wall times are distinct from the component reductions in Table~\ref{tab:lifecycle-results}. The two-worker combination batch completes in 7.49 hours, at 138.18 executions per hour. Batch throughput and per-execution latency measure different effects of concurrency.

\paragraph{Profile reuse amortizes a separate acquisition cost.}
The new JP/AP/JAP batch consumes 109,187,191 fresh executor tokens; reused planning contributes another 1,553,328 attributed tokens, with no new planning or profile calls. Across the batch, failed/rejected calls consume 4,809,883 tokens and non-failing recovery continuations consume 19,308,638. Whole failed executions consume 11,785,992 tokens, overlapping these categories.

The frozen profile draws on 33,942,386 previously purchased JIT+adaptive executor tokens and 1,134,687 structural-profile tokens. Charging this entire 35.077M-token acquisition to reuse gives an amortized cost of $C_{\mathrm{online}}+35.077\mathrm{M}/N$ after $N$ executions. At the observed historical-baseline gap of 75.98K tokens per JAP attempt, the illustrative token-volume break-even is approximately 462 executions. This is a reuse accounting calculation; the acquisition is already paid in the reported runs, and token volume is not a monetary billing model.

\subsection{Per-Task Winners and Selection Headroom}
\label{sec:per-task-winners}

We recompute the per-task winners using the displayed observations: original J/A/JA and the latest P/JP/AP/JAP. For each task, we first retain configurations with the highest success count out of three repetitions, then minimize the all-attempt mean token cost or component time. This success-first rule uses all three attempts, including failures and recovery. Each column in Table~\ref{tab:task-winners} sums to 115, with no exact ties.

\begin{table}[!tbp]
\centering

\setlength{\tabcolsep}{3.5pt}
\begin{tabular}{@{}lrr@{}}
\toprule
Configuration & \shortstack{Token-winning\\tasks} & \shortstack{Latency-winning\\tasks} \\
\midrule
J & 26 & 35 \\
A & 20 & 18 \\
$P^{\dagger}$ & 15 & 15 \\
JA & 22 & 16 \\
$JP^{\dagger}$ & 15 & 15 \\
$AP^{\dagger}$ & 6 & 7 \\
$JAP^{\dagger}$ & 11 & 9 \\
\bottomrule
\end{tabular}
\caption{Success-first winners in the updated seven-configuration display. Highest success count is followed by lowest all-attempt mean cost. $\dagger$ identifies the later PGO evaluations. Each objective assigns one winner to each of 115 tasks.}
\label{tab:task-winners}
\end{table}

\begin{figure*}[!tbp]
\centering
\includegraphics[width=\linewidth,height=0.75\textheight,keepaspectratio]{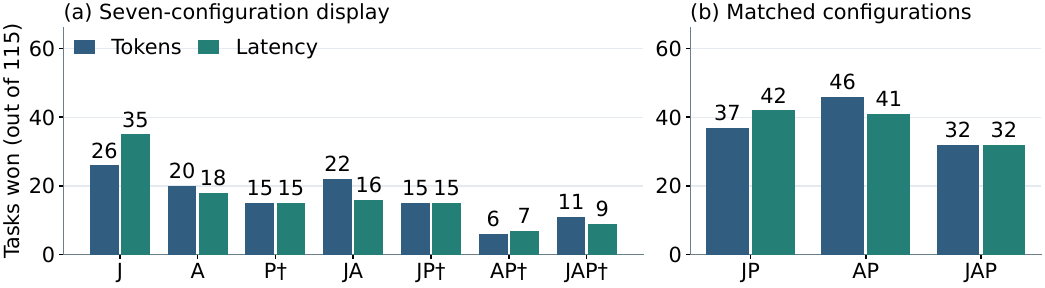}
\caption{Success-first task winners. Left: the updated seven-configuration display across recorded batches. Right: the contemporaneous JP/AP/JAP experiment. Each objective selects one complete configuration per task and retains all three repetitions.}
\label{fig:task-winners}
\end{figure*}

Every configuration wins at least six tasks for tokens and seven for latency in the seven-configuration display. J wins 26 token tasks and 35 latency tasks; P alone wins 15 for each objective. P-containing configurations collectively win 47 token tasks and 46 latency tasks. The distribution supports maintaining multiple lifecycle choices: different tasks expose different combinations of structural, observation, and historical opportunities.

\paragraph{Winning more tasks is not the same as minimizing workload cost.}
Within the matched JP/AP/JAP study, AP wins 46 tasks for tokens, JP wins 37, and JAP wins 32. Nevertheless, JAP has the lowest aggregate token cost. A task count weights a narrow win on a cheap task as much as a large regression on an expensive task; the mean cost retains the size of both. A selector should therefore optimize expected cost at the required completion level, rather than simply predict the most frequent winner label.

\paragraph{A strong default captures much of the observed benefit.}
JAP lies within 5\% of the success-first token-best choice on 85/115 tasks in the matched experiment. A hindsight token oracle that selects one of JP/AP/JAP per task completes 334/345 executions and averages 103.51K tokens, only 0.86\% below fixed JAP. The latency oracle averages 31.21 seconds, 3.26\% below fixed JP. The wider seven-configuration display has more descriptive headroom: its token oracle averages 95.48K tokens, 8.54\% below JAP, with 340/345 successes; its latency oracle averages 29.36 seconds, 8.52\% below P. These hindsight references choose one configuration per task using all observed outcomes, rather than switching independently after each repetition.

Exact winner labels are less stable than competitive sets. Among the 100 matched tasks where every configuration succeeds in all repetitions, the same exact token winner persists across repeats on seven tasks, whereas some configuration remains within 5\% of best on 44 tasks. This favors a policy with a strong default and a meaningful switching margin. Selecting among near-equivalent configurations offers little benefit; identifying expensive recovery paths offers a more substantial opportunity.

\subsection{Scenario-Held-Out Policy Selection}
\label{sec:policy-selection-results}

We evaluate the learned token-oriented selector using five scenario-held-out folds over the original matched lifecycle outcomes; the later P, JP, AP, and JAP runs are not substituted into this evaluation. This selector is distinct from Automatic AgentPRO's heuristic controller in Section~\ref{sec:overall-performance}. Each held-out task receives one configuration choice, applied to its three stored repetitions. Each fold uses only its training data to fit the selector and choose one fixed comparator for its held-out scenarios. Table~\ref{tab:policy-results} includes all failures and recovery costs and places the two hindsight references alongside the held-out policies. These are overlapping views of existing executions, not additional task runs.

\begin{table*}[!tbp]
\centering

\setlength{\tabcolsep}{3.5pt}
\begin{tabular}{@{}lrrrrr@{}}
\toprule
Policy & Success & \shortstack{Mean\\tokens\\($10^3$)} & \shortstack{Token\\saving\\(\%)} & \shortstack{Mean\\time\\(s)} & \shortstack{Time\\reduction\\(\%)} \\
\midrule
Fixed J & 321/345 & 105.42 & 1.70 & 32.45 & 2.26 \\
Fixed JAP & 325/345 & 106.31 & 0.87 & 32.95 & 0.75 \\
Best fixed (training only) & 321/345 & 107.24 & 0.00 & 33.20 & 0.00 \\
Learned selector & 318/345 & 109.90 & -2.47 & 33.77 & -1.71 \\
Token oracle (hindsight) & 339/345 & 98.21 & 8.42 & 30.46 & 8.24 \\
Latency oracle (hindsight) & 339/345 & 108.36 & -1.05 & 29.39 & 11.46 \\
\bottomrule
\end{tabular}
\caption{Offline policy comparison and hindsight references on the original seven-configuration matched outcomes. Percentages use the training-selected best fixed policy as reference; negative values are increases. The learned selector is scenario-held-out. Oracle rows use all outcomes and are optimistic references, not deployable policies. Fixed JAP is a lifecycle mask under the automatic manager, distinct from the earlier always-on policy. All failures and recovery remain included; warm routing cost is included for the selector.}
\label{tab:policy-results}
\end{table*}

\begin{figure*}[!tbp]
\centering
\includegraphics[width=\linewidth,height=0.75\textheight,keepaspectratio]{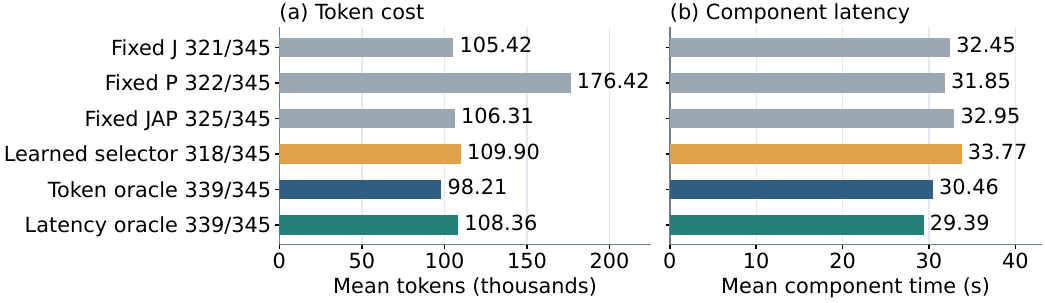}
\caption{Fixed configurations, the held-out learned selector and two objective-specific hindsight oracles on the original seven-configuration matched runs. Counts show official successes. Both resource costs of each oracle's selected configurations are shown; the token and latency minima are not combined into one hypothetical policy.}
\label{fig:oracle-gap}
\end{figure*}

The learned selector uses 2.47\% more mean tokens and 1.71\% more mean component time than the training-selected fixed policy, and completes three fewer executions. Warm routing averages approximately 0.33 milliseconds per decision, excluding artifact loading and offline training, and uses no LLM calls. These measurements locate the remaining policy problem in prediction quality: routing is inexpensive, but a task-level decision must anticipate recovery exposure and the size of a potential saving, not merely identify a different winner.

The original selector results and the refreshed configuration analysis answer complementary questions. The former measures a concrete held-out decision rule; the latter identifies how much useful choice remains and where its cost lies. Together they motivate a strong fixed default, an explicit benefit margin before switching, and features describing live-state requirements and the risk of long recovery paths.

\subsection{Summary of Findings}
\label{sec:results-summary}

The results support three design conclusions. First, Automatic AgentPRO reduces all-attempt mean tokens by 41.85\% and mean component time by 6.61\% relative to the historical baseline, completing 110/115 tasks. Its exercised passes connect recurring prompt and read simplification with changes to generation boundaries. Second, the separate lifecycle study shows that coordination changes recovery exposure and optimization overhead. The latest fixed JAP configuration reduces mean tokens by 42.12\% and component time by 7.28\% relative to the historical baseline, with 331/345 successes. Within the matched JP/AP/JAP batch, JAP gives the best observed token--completion combination, while JP has the lowest component latency. Third, task-dependent winners create selection opportunities, but the learned selector evaluated on the original matched outcomes does not outperform its training-selected fixed comparator. The refreshed matched study places fixed JAP within 5\% of token-best on 85/115 tasks, motivating a strong default and a meaningful expected benefit before switching.

\section{Related Work}
\label{sec:related-work}

\paragraph{Workflow and skill compilation.}
ReWOO separates tool-use reasoning from observation collection to reduce
repeated model input~\cite{rewoo}. LLMCompiler constructs function-call
plans and dispatches independent calls in parallel~\cite{llmcompiler}.
LLM-Tool Compiler fuses compatible tool operations into a single interface
to expose more parallel work per model request~\cite{llm-tool-compiler}.
SkVM compiles skills for model and harness capabilities, binds their
environments, and extracts concurrency; its runtime also supports code
solidification and adaptive recompilation~\cite{skvm}. SkillSmith compiles
skill packages into runtime interfaces that selectively expose relevant
components~\cite{skillsmith}. These systems establish that agent execution
structure can be analyzed and transformed. AgentPRO focuses on the interaction
between generation regions and their operations: a validated operation
rewrite can change a region's observation requirements, while regrouping
can expose further operation rewrites. Its shared IR carries these
requirements across JIT, adaptive optimization, and PGO.

\paragraph{Context and trajectory reduction.}
AgentDiet reduces irrelevant, redundant, and expired trajectory content
during execution~\cite{agentdiet}. ACON compresses observations and history,
improving compression guidelines through agent failure analysis~\cite{acon}.
SkillReducer reduces instruction overhead by compressing skill descriptions
and separating core instructions from references loaded on demand~\cite{skillreducer}.
AgentFold learns to consolidate historical context at multiple
scales~\cite{agentfold}. These methods change the information presented to
the model; several already use online decisions or learned historical
feedback. AgentPRO includes context projection in its operation-pass
library and studies its composition with guarded read reuse and generation
grouping. Reducing the representation of a completed tool response and
eliminating a future redundant tool operation are distinct transformations;
the latter requires checks on arguments, effects, and current result validity.

\paragraph{Reuse of execution history.}
Agentic Plan Caching extracts reusable plan templates from completed
executions and adapts them to subsequent requests~\cite{plan-caching}.
TraceCompiler mines traces into mostly deterministic workflows, using
evidence-backed dependencies and retaining model decisions where compilation
is unsupported~\cite{tracecompiler}. PAIR uses matched continuations from
the same agent state to identify harmful compression events and revise
compression prompts~\cite{pair}. These approaches demonstrate several uses
of execution history: reusing procedures, compiling recurring behavior,
and improving transformation policies. AgentPRO's PGO uses history to
guide region and operation passes, including grouping and promotion of
recurring adaptive rewrites. Current checks govern the applicability of
these policies. Our evaluated setting generates fresh executor code rather
than replaying a previous program or answer; historical profiles guide the
workflow that produces that code.

\paragraph{Optimization policy and net cost.}
StateComp learns when historical interactions can be replaced, using the
current agent state to select spans for compression~\cite{statecomp}.
The Complexity Trap compares observation masking, summarization, and their
combination, finding that additional compression complexity does not
necessarily improve execution efficiency~\cite{complexity-trap}.
Beyond Token Savings separately studies compression target, timing, and
strength, showing that token reduction need not imply lower latency and
that effects vary across tasks and models~\cite{beyond-token-savings}.
These findings motivate evaluating transformations by their downstream
execution costs and success, rather than prompt size alone. AgentPRO makes
JIT, adaptive optimization, and PGO independently selectable and records
optimization overhead, failed executions, and recovery costs. Its evaluation
separates configuration effectiveness from policy-selection quality;
task-dependent winners alone do not establish that a learned selector can
outperform a fixed configuration.

\label{page:references-start}
\bibliographystyle{ACM-Reference-Format}
\bibliography{ref}

\end{document}